\documentclass[cpp,a4paper,fleqn]{w-art}

\usepackage{times,cite,w-thm}
\theoremstyle{plain}

\theoremstyle{definition}

\usepackage[]{graphicx}

\usepackage{subfigure}
\usepackage[english]{babel}
\newcommand{\beq}{\begin{equation}}
\newcommand{\eeq}{\end{equation}}
\newcommand{\bea}{\begin{eqnarray}}
\newcommand{\eea}{\end{eqnarray}}

\begin{document}

\DOIsuffix{theDOIsuffix}

\Volume{XX}
\Month{XX}
\Year{XXXX}

\pagespan{1}{}

\Receiveddate{XXXX}
\Reviseddate{XXXX}
\Accepteddate{XXXX}
\Dateposted{XXXX}


\title[Quark-gluon plasma thermodynamics at finite chemical potential. Color PIMC simulations]{Thermodynamics of the quark-gluon plasma at finite chemical potential: color path integral Monte Carlo results
  }%
\author[V. Filinov {\em et al.}]{V.S.~Filinov\inst{1}
  \footnote{Corresponding author\quad E-mail:~\textsf{vladimir\_filinov@mail.ru},
}}
\author[]{M. Bonitz\inst{2}}
\author[]{Y.B.~Ivanov\inst{3}}
\author[]{M.~Ilgenfritz\inst{4}}
\author[]{V.E. Fortov\inst{1}}
\address[\inst{1}]{Joint Institute for High Temperatures, Russian Academy of
Sciences, Izhorskaya 13, bd. 2, 125412 Moscow, Russia}
\address[\inst{2}]{Institute for Theoretical Physics and Astrophysics, Christian Albrechts University Kiel, \\ Leibnizstrasse 15, D-24098 Kiel, Germany}
\address[\inst{3}]{Russian Research Center ``Kurchatov Institute'', Kurchatov Sq. 1,  123182 Moscow, Russia}
\address[\inst{4}]{Joint Institute for Nuclear Reseach, Joliot-Curie str., 6, Dubna, Moscow Region,
141980, Russia}
\date{\today}
\begin{abstract}
Based on the constituent quasiparticle model of the quark-gluon 
plasma (QGP), color quantum path-integral Monte-Carlo (PIMC) calculations of the thermodynamic 
properties of the QGP are performed. We extend our previous zero chemical potential simulations to the QGP at 
finite baryon chemical potential. The results indicate that color PIMC can be applied not only above the
QCD critical temperature $T_c$ but also below $T_c$. Besides reproducing the lattice equation of state our approach yields 
also valuable additional insight into the internal structure of the QGP, via the pair distribution 
functions of the various quasiparticles. In particular, the pair distribution function of gluons
reflects the existence of gluon-gluon bound states at low temperatures and $\mu=175$ MeV, i.e. glueballs, 
while meson-like bound states are not found. 
\end{abstract}

\maketitle

\section{Basics of the QGP model and comparison with PIMC for plasmas}\label{semi:model}
Strongly correlated charged particle systems have attracted growing interest over the recent three decades in many fields. This includes laser compressed plasmas \cite{glenzer_sccs11}, ions in traps, dusty plasmas \cite{bonitz_rpp10} or dense plasmas in planet cores. For the theoretical description of dense quantum plasmas 
path integral Monte Carlo (PIMC) simulations have proved particularly successful, e.g. \cite{mil-pol,filinov_ppcf01}. 
A few years ago experiments at the Relativistic  Heavy-Ion Collider (RHIC) at Brookhaven National Laboratory 
 \cite{Shuryak0} and at the Large Hadron Collider (LHC) at CERN  \cite{Schukraft} have produced an unconfined quark-- gluon plasma (QGP) which turned out to 
behave as a nonideal liquid \cite{Shuryak0,Shen},  
Although equilibrium properties of the strongly QGP are successfully computed using lattice QCD \cite{Shuryak0,Borsanyi10,Cheng}, these simulations are 
very time consuming and not easy to interpret. Also, they fail, e.g. at large quark chemical potential. Based on the above mentioned experience with PIMC simulations of strongly correlated Coulomb systems it is, therefore, tempting to make these methods available also for the description of the QGP. We have developed such ``color PIMC'' simulations in recent years, e.g. \cite{filinov_cpp11,Filinov:2012zz,FilPRV} focusing on zero baryon chemical potential. Here we extend these simulations to the important case of finite chemical potential. In the following we briefly discuss the model \cite{filinov_ppcf14} and present first results for the thermodynamic properties.

To start with we provide a comparison of an (``electromagnetic'') electron-ion plasma and a quark gluon plasma, see Table \ref{tab:comparison}, since this provides the basis to understand the main physical ingredients required for realistic color PIMC simulations. Although QCD was constructed in analogy to quantum electrodynamics there exist fundamental differences. While Coulomb interacting charges are mapped on fermions (or bosons) whose interaction is mediated by (usually weakly interacting) photons the situation in QCD is different. Here also the field particles (gluons) providing the interaction between fermions (quarks and antiquarks) are, in general, strongly interacting. It is, therefore, advantageous to consider not bare quarks, antiquarks and gluons but quasiparticles by abosrbing the ``hard modes'' into the quasiparticles
whereas the soft modes are incorporated into an effective color Coulomb interaction \cite{filinov_ppcf14}.

\begin{table}
\caption{Comparison of an Electromagnetic plasmas and the QGP model simulated in this paper. {\bf Comments and notations}: $T$: temperature, $\mu$: chemical potential,  $r_{ab}$: distance between particles a and b, QP: quasiparticles. $^a$ Only intrinsic, no orbital quantum numbers are listed. $^b$ defined by the kinetic energy operator.}
\begin{center}
\begin{tabular}{|l||c|c|}
\hline
              & {\bf Electromagnetic plasma} & {\bf Semiclassical SU(3) quark-gluon plasma} \\
\hline\hline
Basic particles/QP      &electrons, ions (i) or holes (h) & quarks, antiquarks, gluons \\
\hline
Quantum numbers$^a$  & spin  & spin, flavor, color \\
\hline
Renormalization &  none (plasmas), QP in solids  & QP \\ 
\hline 
Masses & $m_e \ll m_i$ or QP masses $m_e \sim m_h$ & comparable QP masses, $m_\alpha=m_\alpha(T,\mu)$ \\
\hline
Charge                   & fixed scalar electrical charge $q_\alpha$ & SU(3) Wong vector color charge variables $Q_\alpha$ \\
\hline
Coupling constant & fixed value, $\alpha = 1/137$ & state and distatnt dependent, $\alpha (T,\mu, r_{ab}) $\\
\hline
Potential energy         &non-relativistic Coulomb potential & non-relativistic color  Coulomb potential \\
                         & $V_{ab} \sim q_a q_b/r_{ab}$ & $V_{ab} \sim \alpha (T,\mu, r_{ab}) Q_a \bullet Q_b/r_{ab}$ \\
                         &   & $Q_a \bullet Q_b$ scalar product of 8D vectors \\
\hline
Kinetic energy  & non-relativistic & relativistic \\
\hline
Path integral & non-relativistic & relativistic Bessel and \\
partition function$^b$  & Gaussian measure & SU(3) group Haar measures \\
\hline
\end{tabular}
\end{center}
\label{tab:comparison}
\end{table}

The basic assumptions of our model are similar to those of  Ref. \cite{shuryak1}:

\begin{description}
 \item[I.]
 Quasiparticles masses ($m$) are of order or higher
 than the mean kinetic energy  per particle. This assumption is based on the analysis of QCD lattice
 data \cite{Lattice02,LiaoShuryak,Karsch:2009tp}. For instance, at zero net-baryon density it amounts to $m \sim  T$, where $T$ is a temperature.
 \item[II.] 
 In view of I., the interparticle interaction is dominated by a color-electric Coulomb potential.
  \item[III.] Since the color representations are large, the color operators 
 are substituted by their average values [by Wong's classical color vectors, i.e. eight-dimensional (8D) vectors in SU(3)] with the quadratic and cubic Casimir conditions \cite{Wong}.
 \item[IV.] We consider the 3-flavor ('$u$p', '$d$own' and '$s$trange') quark model assuming equal quark masses. 
 The gluon (quasiparticle) mass is allowed to be different (heavier) from that of the quarks.
\end{description}
Thus, this model requires the following three input quantites as a function of temperature ($T$)
and quark chemical potential ($\mu_q$):
the quasiparticle masses, for quarks $m_q$ and gluons $m_g$, and
the coupling constant $g^2$, or $\alpha_s = g^2/4\pi$.
These input quantities should be deduced from lattice QCD data or from 
an appropriate model simulating these data. 

The applicability of such an approach was discussed in Refs. \cite{LM105,shuryak1} in detail.
It has been established that hard modes (in terms of {\em hard thermal loop} approximation
\cite{Pisarski89,Braaten90,Blaizot02}) behave like quasiparticles. 
Therefore, masses of these quasiparticles should be deduced from 
nonperturbative calculations taking into account hard field modes,
e.g., they can be associated with pole masses deduced
from lattice QCD calculations.
At the same time, the soft quantum
fields are characterized by very high occupation numbers per mode. Therefore,
to leading order, they can be well approximated by soft classical fields.
This is precisely the picture we are going to utilize: massive
quantum quasiparticles (hard modes) interacting via classical color fields.
 Our approach
differs from that of Ref. \cite{shuryak1} by a quantum treatment to quasiparticles instead
of the classical one, and additionally by a relativistic desciption of the kinetic energy instead of the nonrelativistic
approximation of Ref. \cite{shuryak1}.

\section{Color Path Integrals}\label{s:pimc}
We consider a multi-component QGP consisting of $N$ color quasiparticles:  $N_g$  gluons,  $N_q$  quarks and $\overline{N}_{q}$ antiquarks. The Hamiltonian of this system is
 $\hat{H}=\hat{K}+\hat{U}^C$, consisting of kinetic and color Coulomb interaction parts
\begin{eqnarray}
\label{Coulomb}
{\hat{K}}&=&\sum_i \sqrt{\hat{\bf p}^2_i+m^2_i(T,\mu_q)},
\qquad
{\hat{U}^C} = \frac{1}{2}\sum_{i\neq j}
\frac{g^2(T,\mu_q)
(Q_i\cdot Q_j)}
{4\pi| {\bf r}_i- {\bf r}_j|}.
\end{eqnarray}
Here the $i$ and $j$ summations run over all quark and gluon quasiparticles, $i,j=1,\ldots,N$, with
$N=N_q+\overline{N}_q+N_g$, where 
$N_q=N_u+N_d+N_s$ and $\overline{N}_{q}=\overline{N}_{u}+\overline{N}_{d}+\overline{N}_{s}$ are the total
numbers of quarks and antiquarks of all included flavours ({\em u}p, {\em d}own and {\em s}trange), respectively.
Further, ${\bf r}_i$ are 3D vectors  of the quasiparticle spatial coordinates, whereas $Q_i$ denote
 Wong's quasiparticle color variable [8D-vectors in the $SU(3)$ group], and
$(Q_i\cdot Q_j)$ denotes the scalar product of two color vectors.
 For the potential energy the non-relativistic approximation is used whereas
 for the kinetic energy the full relativistic form is retained, since the quasiparticle masses are not negligible
 compared to temperature. The eigenvalue equation of this Hamiltonian (the ``spinless Salpeter equation'') may be regarded as a 
well-defined approximation to the Bethe-�Salpeter formalism
 \cite{SB1,SB2,LS1,LS2} for the description of bound states within relativistic quantum field theory. 

Thermodynamic properties in the grand canonical ensemble with a given inverse temperature $\beta=1/T$,
net-quark and strange chemical potentials $\mu_q$ and $\mu_s$ and fixed volume $V$ are fully described by the
grand partition function
%
%
\begin{eqnarray}
Z\left(\mu_q,\mu_s,\beta,V\right)
& = &
\sum_{\{N\}}
\frac{ e^{\beta\mu_q(N_q-\overline{N}_{q})}\;e^{\beta\mu_s(N_s-\overline{N}_{s})}}%
{N_u!\;N_d!\;N_s!\;\overline{N}_{u}!\;\overline{N}_{d}!\;\overline{N}_{s}!\;N_g!}
Z\left(\{N\},V,\beta\right) ,
\label{Gq-def}
\\
Z\left(\{N\},V,\beta\right) &=& 
\sum_{\sigma} \int\limits_V dr\; d\mu Q \;\rho(r,Q, \sigma; \{N\}; \beta),
\label{Z-def}
\end{eqnarray}
where the variables 
$\{N\}=\{N_u,N_d,N_s,\overline{N}_{u},\overline{N}_{d},\overline{N}_{s},N_g\}$ 
are independent and 
$\rho(r,Q, \sigma;\{N\} ; \beta)$ denotes the diagonal matrix
elements of the density  operator ${\hat \rho} = \exp (- \beta{\hat H})$.
Here $r$, $\sigma$ and  $Q$  denote the multi-dimensional vectors related to spatial, spin and color
degrees of freedom, respectively, of all quarks, antiquarks and gluons. 
The $\sigma$ summation and the spatial ($dr\equiv d^3 r_1 ...d^3 r_N $)
and color ($d\mu Q\equiv d\mu Q_1 ... d\mu Q_N $) integrations
run over all individual degrees of freedom of the quasiparticles,
whereas $d\mu Q_i$ denotes integration over the SU(3) group Haar measure \cite{LM105}.
In Eq. (\ref{Gq-def}) we explicitly wrote a sum over different quark flavors (u,d,s) 
and assume that the strange chemical potential, $\mu_s=-\mu_q$ (nonstrange matter). 
Since the masses and the coupling constant depend on temperature and on the quark chemical potential,
all thermodynamic functions 
should be calculated through the respective derivatives of the logarithm of the partition function.
%

The (unknown) exact density operator ${\hat \rho} = e^{-\beta {\hat H}}$ of interacting quantum
systems can be constructed using a path integral
approach
based on the operator identity
$e^{-\beta \hat{H}}= e^{-\Delta \beta {\hat H}} \cdot
e^{-\Delta \beta {\hat H}}\dots  e^{-\Delta \beta {\hat H}}$,
where the r.h.s. contains $n+1$ identical factors  with $\Delta \beta = \beta/(n+1)$.
The main advantage of this identity is that it
allows us to use perturbation theory to obtain an approximation for each of the factors,  
which is applicable due to the smallness{\em ?? large value } of the temperature $1/\Delta \beta$. Each factor 
should be calculated with an accuracy of the order of $1/(n+1)^\theta$, 
with $\theta > 1 $, as in this case the error of the whole product, in the limit of $n \to \infty$, 
will be equal to zero. 
Generalizing the electrodynamic plasma results \cite{filinov_ppcf01} to the quark-gluon plasma case 
(see table~\ref{tab:comparison}),
we may use the approximation\footnote{For the sake of notational convenience, we ascribe the superscript $^{(0)}$
to the original variables.}
%
\begin{eqnarray}
&&
\rho = 
e^{-\beta U}
 \,
 \frac{{\rm per}\,||\widetilde{\underline{\phi}}^{(n),(0)}||_{N_g}}{\tilde{\lambda}_g^{3N_g}}
 \,
\frac{{\det}\,||\tilde{\phi}^{(n),(0)}||_{N_q} \,}{\tilde{\lambda}_q^{3N_q}}
\,
\frac{{\det}\,||\tilde{\phi}^{(n),(0)}||_{N_q} \,}{\tilde{\lambda}_{{\bar{q}}}^{3\bar{N}_q}}
\,
\prod\limits_{l=1}^n \prod\limits_{i=1}^N
\phi^{(l)}_{ii} \, , \label{Grho_s}
\end{eqnarray}
%
where the proper (anti-)symmetrization for  gluons (quarks/antiquarks) 
 results in the permanents (determinants) in Eq.~(\ref{Grho_s}) while 
 the effective total color Coulomb interaction energy 
\begin{eqnarray}
U = 
\frac{1}{n+1}\sum_{l=1}^{n+1} \frac{1}{2}
\sum_{i,j (i\neq j)}^N 
\frac{\Delta\beta}{2}  \left[\Phi^{ij}\left(x_{ij}^{(l-1)},x_{ij}^{(l-1)},Q_i^{(0)},Q_j^{(0)}\right) + 
 \Phi^{ij}\left(x_{ij}^{(l)},x_{ij}^{(l)},Q_i^{(0)},Q_j^{(0)}\right)\right]
\label{up}
\end{eqnarray}
is described in terms of the 
two-particle effective quantum Kelbg color potential  \cite{filinov_ppcf01,Filinov:2012zz} 
\begin{eqnarray}
\Phi^{ij}\left( x_{ij}^{(l)},x_{ij}^{(l)},Q_i,Q_j,\Delta\beta\right) &=&
 \frac{g^2(T,\mu_q)\,\langle Q_i|Q_j \rangle}{4 \pi 
| r_{i}^{(l)}-r_{j}^{(l)}|} \,\Phi\left(x_{ij}^{(l)}\right),
\end{eqnarray}
with $x_{ij}^{(l)}=| r_{i}^{(l)}-r_{j}^{(l)}|/\Delta\lambda_{ij}$,  
$\Delta\lambda_{ij}=\sqrt{2\pi\Delta\beta /m_{ij}}$, $m_{ij}=m_{i}m_{j}/(m_{i}+m_{j})$ and 
$\Phi(x) = 1-e^{-x^2} + \sqrt{\pi} x \left[1-{\rm erf}(x)\right] $ . 
The remaining quantities in  Eq.~(\ref{Grho_s}) are defined as follows: 
$\widetilde{\lambda}^3_a=\lambda^3_a\sqrt{0.5\pi/(\beta m)^5}$ with
  $\lambda_a=\sqrt{2 \pi \beta / m_a}$
 being the thermal  wavelength of a quasiparticle of type ``a'' ($a=q, \bar{q},g$).

Here we have introduced the functions $\phi^{(l)}_{ii}\equiv K_2(z_i^{(l)})/(z^{(l)})^2_i$, 
involving modified Bessel functions $K_2$, where $z_i^{(l)}=\Delta\beta m_i(T,\mu_q)\sqrt{1+
  \left|\xi^{(l)}_i\right|^2 /\Delta\beta^2}$.    
The gluon matrix elements in the permanents are  
$\underline{\phi}_{i,j}^{(n),(0)}=K_2(z_{i,j}^{(n),(0)})/(z_{i,j}^{(n),(0)})^2\delta_\epsilon(Q^{(0)}_i-Q^{(0)}_j)$, 
 while the quark and antiquark  matrix elements entering the determinants are given by
$\tilde{\phi}_{i,j}^{(n),(0)}=K_2(z_{i,j}^{(n),(0)})/(z_{i,j}^{(n),(0)})^2\delta_\epsilon(Q^{(0)}_i-Q^{(0)}_j)\delta_{f_i,f_j}\delta_{\sigma_i,\sigma_j}$
and depend additionally on the spin variables $\sigma_i$ and the flavor index $f_i$ of the particle, which can take the values ``up'', ``down'' and ``strange''. 
 $\delta_\epsilon$ is a broadened delta function of the color vectors whereas the Kronecker symbols properly restrict Pauli blocking to particles with 
the identical spins and flavors. Also, we denoted 
$z_{i,j}^{(n),(0)}=\Delta\beta m_i(T,\mu_q) \sqrt{1+\left|r_{i}^{(0)}-r_{j}^{(n)}
\right|^2/\Delta\beta^2}$. 
The coordinates of the quasiparticle ''beads''  $r_{i}^{(l)} = r_{i}^{(0)}+y_{i}^{(l)}$, ($l>0$) are expressed in terms of $r_{i}^{(0)}$ and 
vectors between
neighboring beads of an $i$ particle, defined as
$y_i^{(l)}= \sum_{k=1}^{l}\xi_i^{(k)}$,  
while  $\xi_{i}^{(1)}, \dots , \xi_{i}^{(n)}$ are 
vector variable of integration  at multiplication of the coordinate representation of mentioned 
above operator identity. 

In the limit $n\rightarrow\infty$ the functions $\phi^{(l)}_{ii}$ describe the {\em relativistic measure} of the color path integrals \cite{FilPRV, filinov_ppcf14}. This measure is created by the {\em relativistic kinetic energy operator} $K=\sqrt{p^2+m^2(T,\mu_q)}$ and, in the limit of large particle mass, coincides with the Gaussian measure used in Feynman's and Wiener's path integrals (see table~\ref{tab:comparison}). 

\section{Color PIMC thermodynamic simulations of the QGP}\label{s:model}
\begin{figure}[htb]
\vspace{0cm} \hspace{0.0cm}
\includegraphics[width=7.0cm,clip=true]{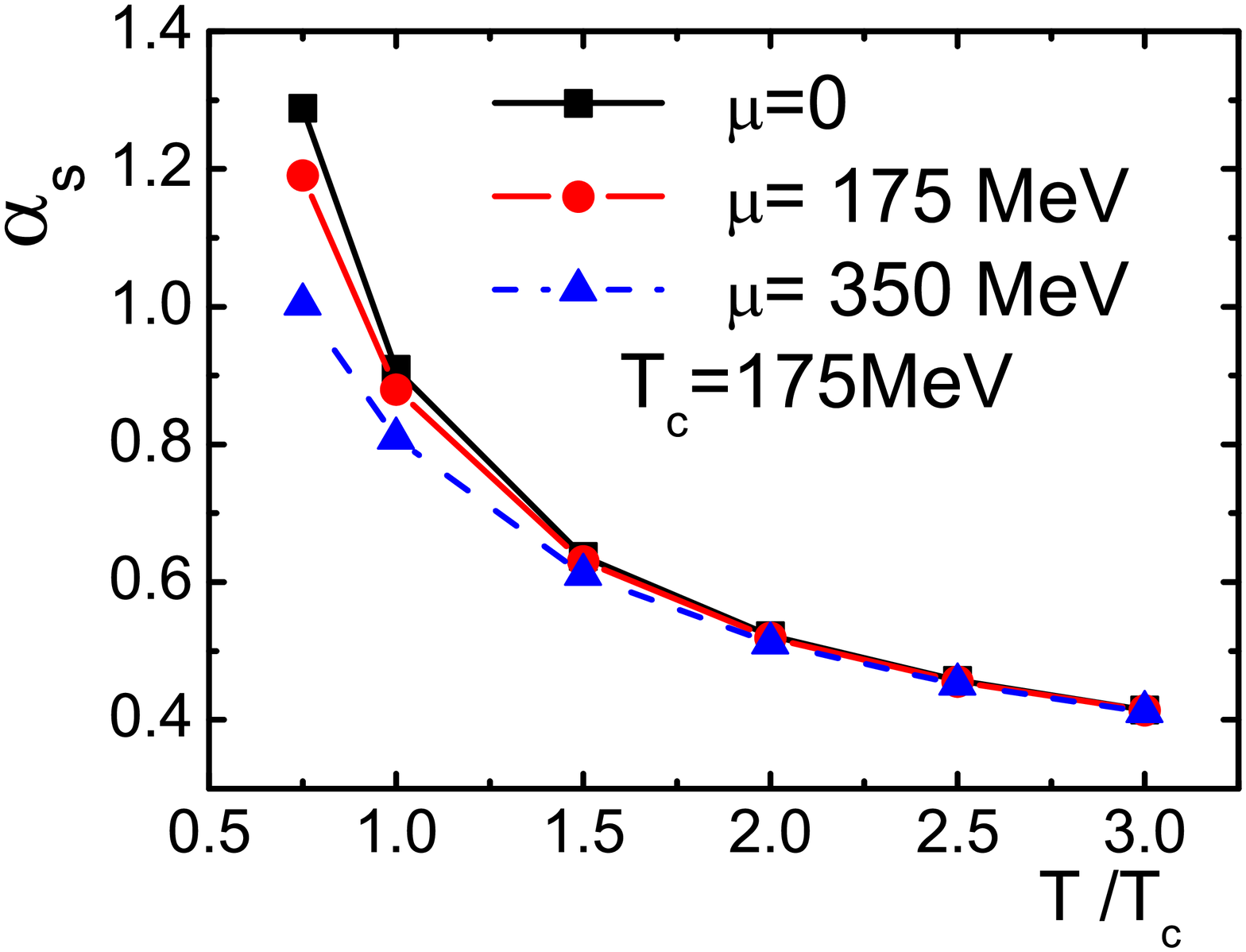}
\includegraphics[width=7.0cm,clip=true]{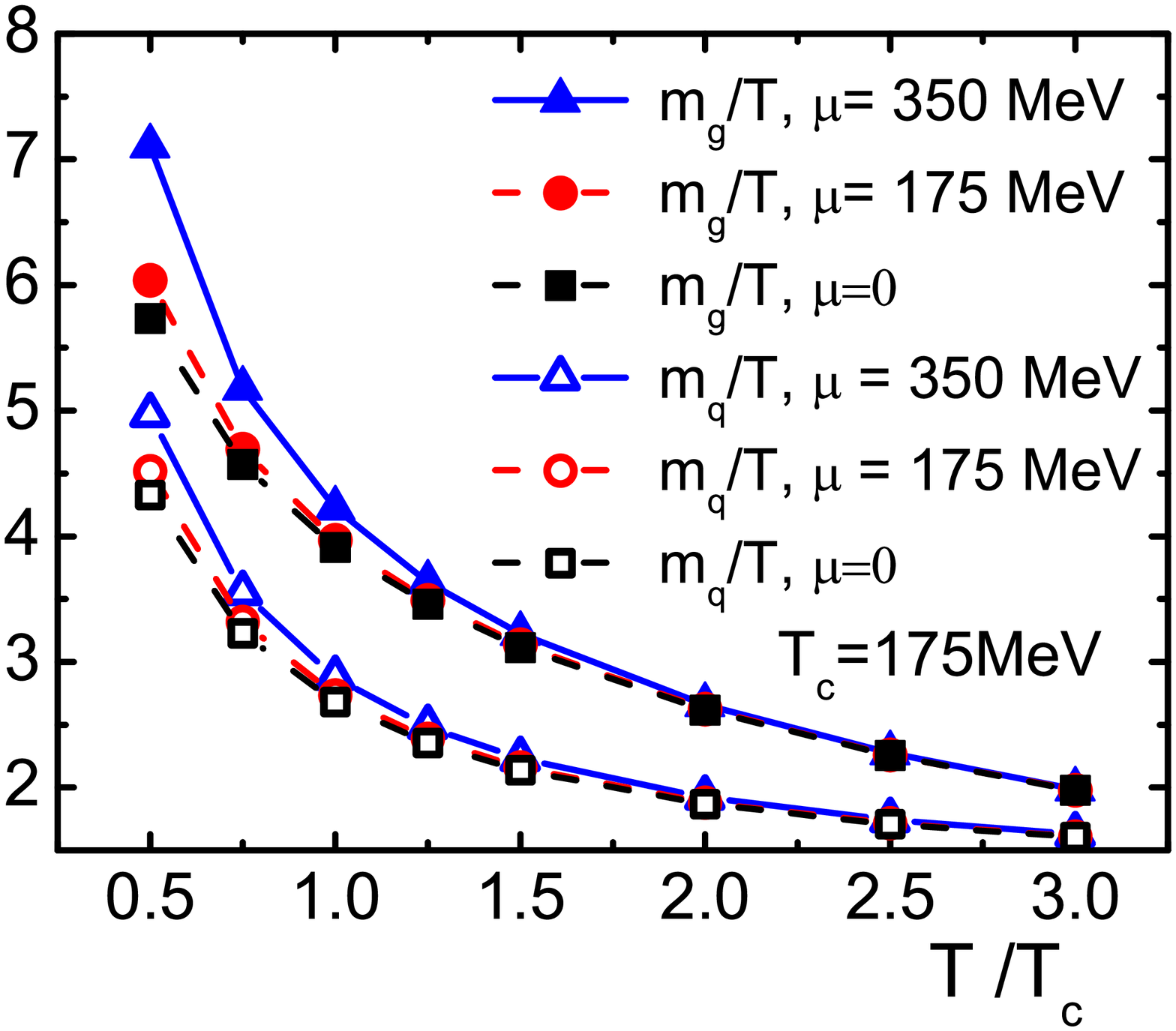}
\caption{
Temperature dependences of the model input quantities.
{\bf Left:} coupling constant $\alpha_s$.
{\bf Right:} quasiparticle mass-to-temperature ratio ($m_q=m_{\bar{q}}$). 
}
\label{fig:alfrs}
\end {figure}

Ideally the parameters of the model should be deduced from the QCD lattice data. However, presently this task is still quite ambiguous. Therefore, in the present simulations we take only one possible set of parameters and extend the analytical statements that are known for high temperatures to the lower temperatures of interest for the present analysis. The HTL perturbative values of $m_g$ and $m_q$  are  known and given for $T\gg T_c$ by \cite{Lebel96} 
\begin{eqnarray}
\label{m_g-qg-T}
m_g^2 (\{\mu_{q}\},T) &=&
\frac{1}{12}\left( (2N_c+N_f)T^2 +
\frac{3}{\pi^2}\sum_{q=u,d,s}
\mu_{q}^2\right)g^2(\{\mu_{q}\}),
\\
\label{m_q-qg-T}
m_{q}^2(\{\mu_{q}\},T) &=&
\frac{N_g}{16N_c}\left(T^2+\frac{\mu_{q}^2}{\pi^2}\right)g^2(\{\mu_{q}\}), 
\end{eqnarray}
where  $N_f$
is the number of quark flavors that can be excited, $N_c=3$ for the SU(3) group, and
$g$ is the QCD running coupling constant, generally
depending on $T$ and all $\mu_q$. According to Eqs.~(\ref{m_g-qg-T}, \ref{m_q-qg-T}) the masses do not depend on 
$T$ and $\mu_{q}$ separately but on their combinations
$
z_{g}=\left( T^2 +
\frac{3}{\pi^2 (2N_c+N_f)}\sum_{q=u,d,s}
\mu_{q}^2\right)^{1/2} 
$
and 
$
z_{q}=\left(T^2+\frac{\mu_{q}^2}{\pi^2}\right)^{1/2}
$, respectively.
It is also reasonable to assume that $g^2$ is a function of this single variable $z_{g}$.
This choice is done because $g^2$ (like the gluons) is related to the whole system rather than to one specific
quark flavor. Then we can use the same ``one-loop analytic coupling'' \cite{Shirkov,Prosperi} 
\bea
\alpha_s (Q^2) = \frac{4\pi}{11-(2/3)N_f} \left[
\frac{1}{\ln (Q^2/\Lambda_{QCD}^2)} + \frac{\Lambda_{QCD}^2}{\Lambda_{QCD}^2-Q^2}
 \right],
\eea
and use in our simulations $2\pi z_g$ for $Q$. This coupling constant $\alpha_s(z_{g}) = g^2/(4\pi)$ 
is displayed in the left panel of Fig.~\ref{fig:alfrs}.  
To obtain model input formulas for the masses $m_g$ and $m_q$ we use 
the related dependencies from our paper \cite{FilPRV} where we replace 
the temperature $T$ by the expressions for $z_{g}$ ($z_{q}$) written above. The final dependencies 
are presented by Fig.~\ref{fig:alfrs} (right panel). 
%
\begin{figure}[htb]
\vspace{-0.5cm} 
\includegraphics[width=7.0cm,clip=true]{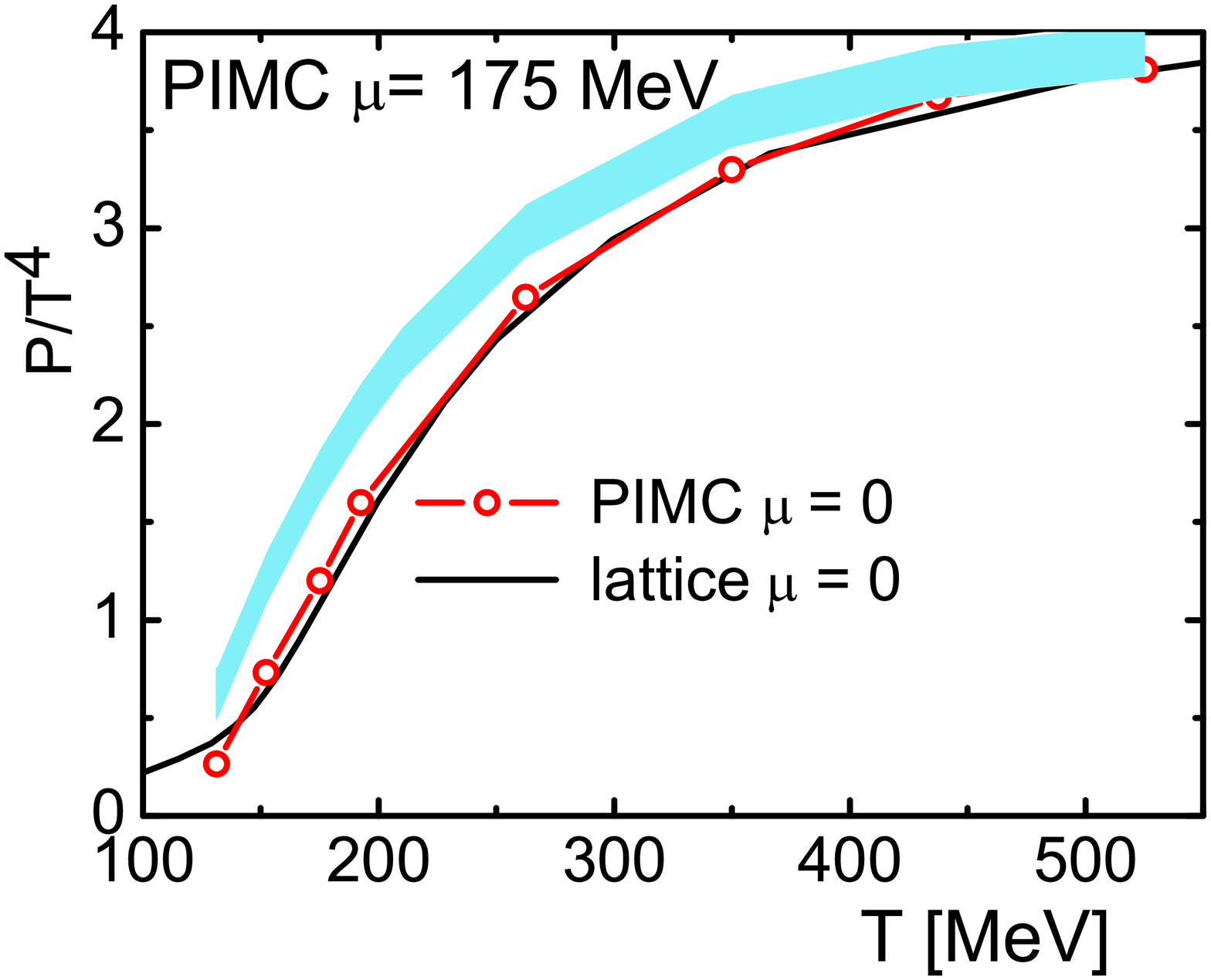}
\vspace{-0.5cm} \hspace{-1.3cm}
\includegraphics[width=7.0cm,clip=true]{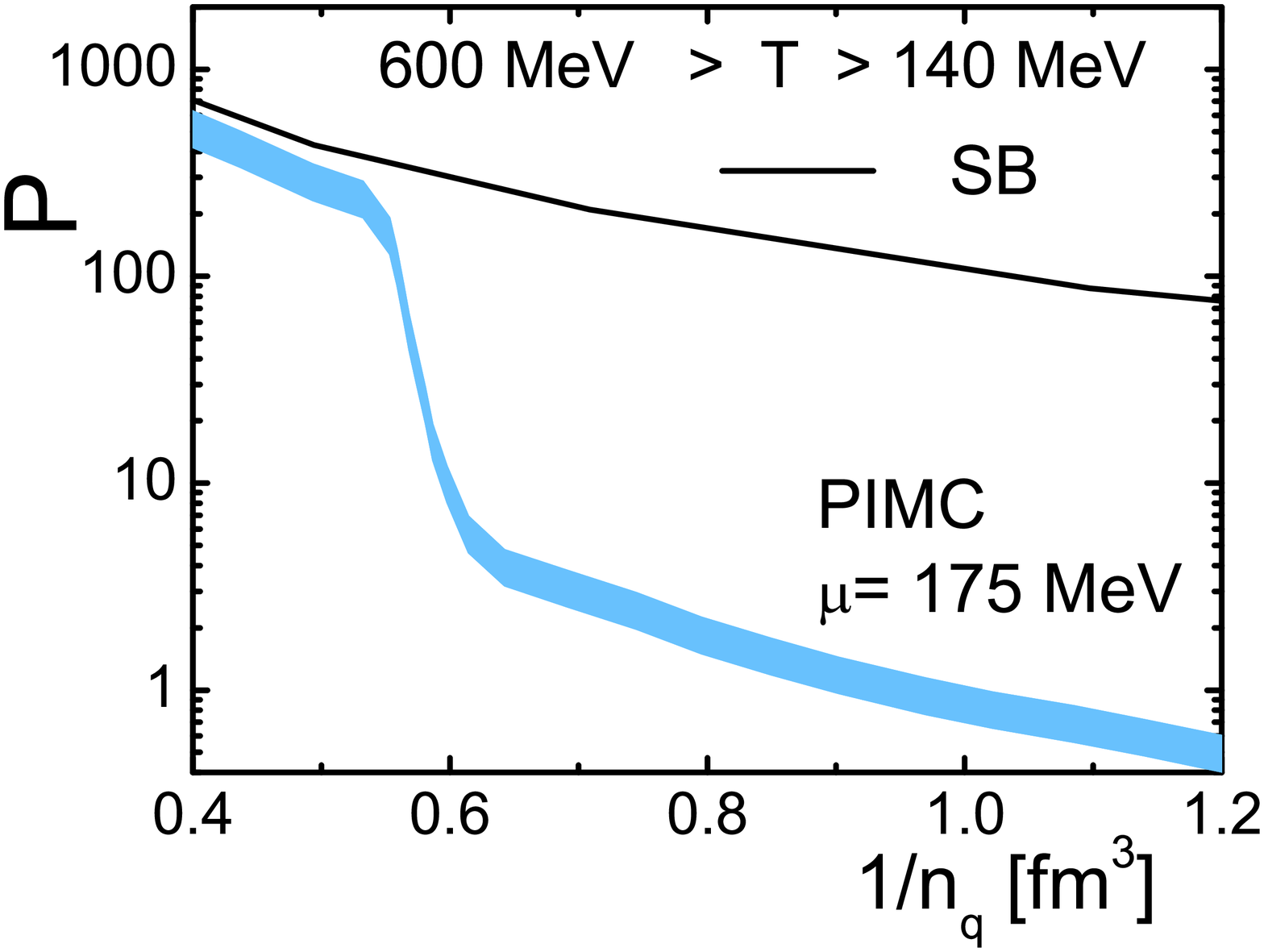}
\includegraphics[width=7.0cm,clip=true]{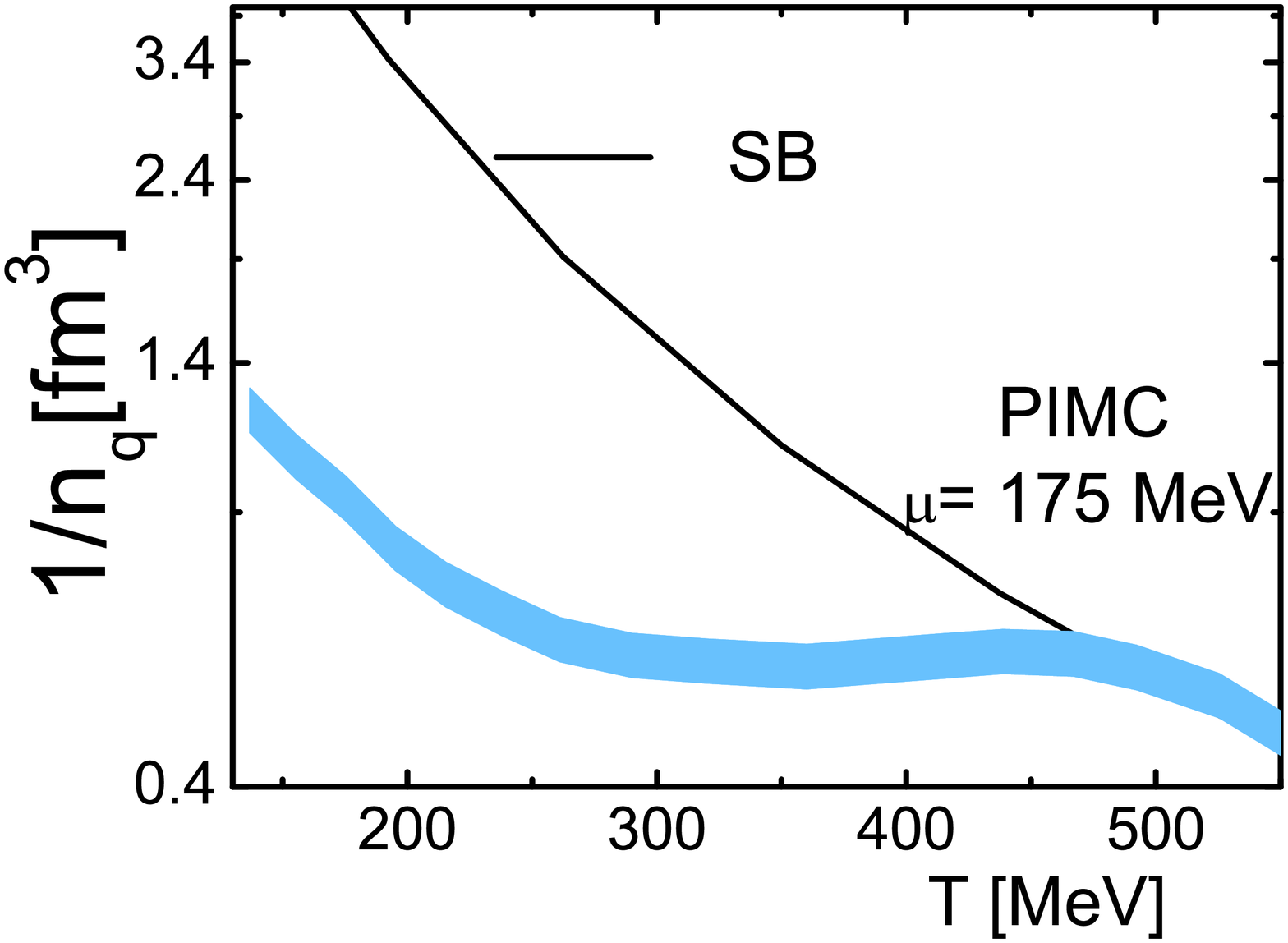}
\vspace{0cm} \hspace{1.7cm}
\includegraphics[width=7.0cm,clip=true]{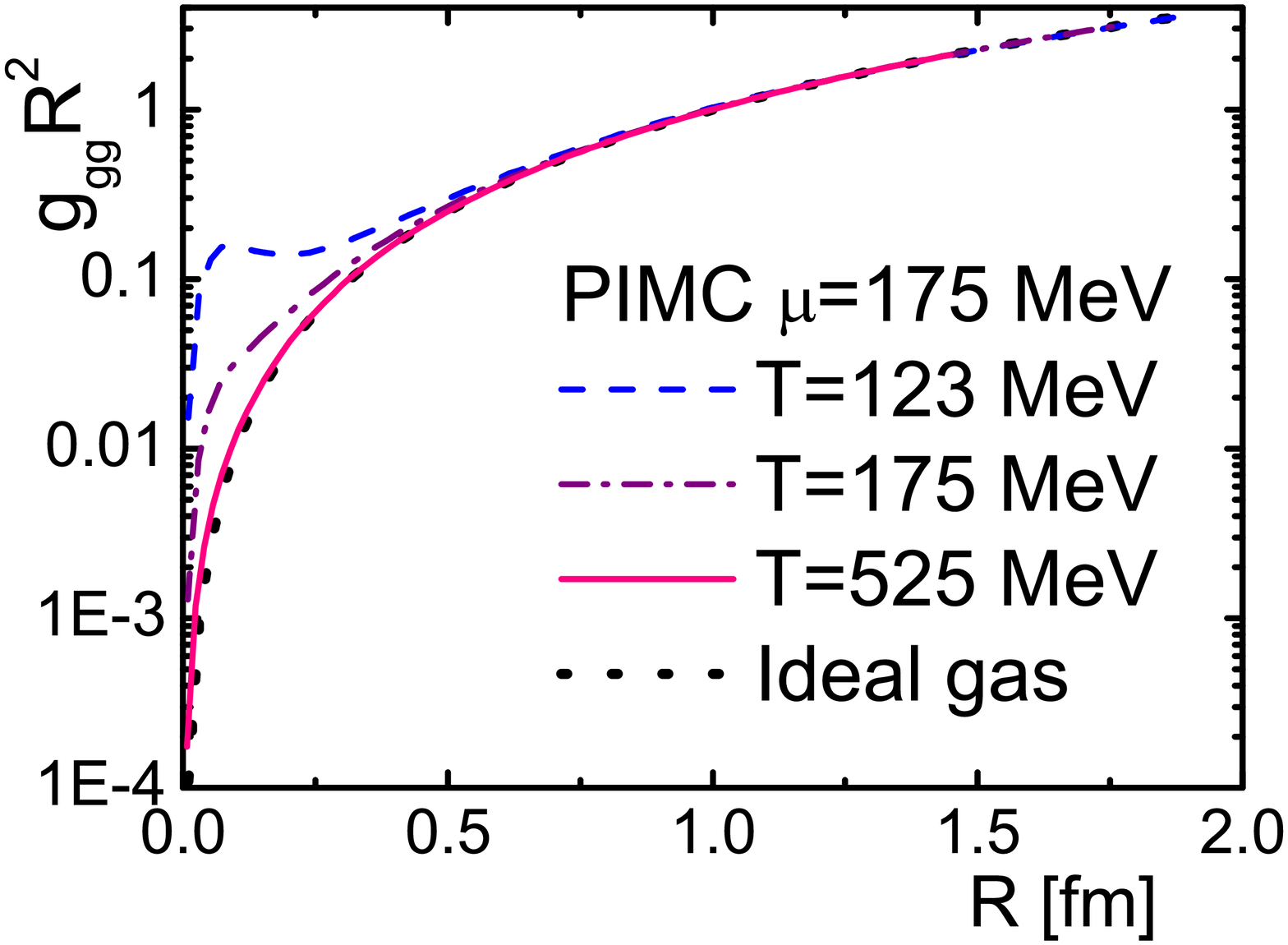}
\vspace{0cm} \hspace{0.0cm}
\caption{Thermodynamic properties of the quark-gluon plasma,
{\bf Top left:} Color PIMC (symbols) and lattice QCD (black line) \cite{Borsanyi10} equation of state of 
 for $\mu=0$, compared to the present color PIMC results for $\mu=175$ MeV (blue area). {\bf Top right}: Stefan--Boltzmann (SB, black line) and PIMC 
 pressure (blue) versus inverse quark density for $\mu=175$ MeV.
 {\bf Bottom left}: Inverse quark density versus temperature for $\mu=175$ MeV -- comparison of Stefan--Boltzmann (SB, black line) limit and the present grand canonical color PIMC results (blue). {\bf Bottom right}: 
 PIMC gluon--gluon quasiparticle pair distributions for different temperatures at $\mu=175$ MeV.  
}
\label{fig:EOS}
\end {figure} 

From Fig.~\ref{fig:EOS} (upper left panel) it follows that the lattice QCD and color PIMC 
equations of state (EOS) of the QGP agree well for zero chemical potential. Similarly, good agreement of
the lattice QCD Taylor expansion technique \cite{Borsanyi12} and color PIMC EOS (blue area for $\mu=175$ MeV) is observed
at non-zero chemical potential \cite{filinov_ppcf14}.
Interesting is also the behavior of pressure versus inverse density (specific volume, top right part of Fig.~\ref{fig:EOS}) and of the inverse density 
 versus temperature (bottom left panel). At high temperatures our color PIMC results are close to the ideal Stefan--Boltzmann. However, at lower temperature, 
the influence of interactions is growing and results in a sharp pressure decrease whereas the quark density changes only weakly. An analogous behavior is observed in  
 a dense electron-ion plasma and is connected with the formation of atomic and molecular bound states \cite{filinov_ppcf01, filinov_pla00}. The same is observed in electron--hole plasmas in semiconductors (formation of excitons and bi-excitons \cite{Fehske,schleede}). The physical origin of the analogous behavior in the present QGP is, thus, the strong quasiparticle interaction which may give rise to the formation of bound states. In fact, we observe evidence of gluon--gluon bound states and gluon balls at low temperatures, which can be seen  
in our gluon--gluon quasiparticle  pair distribution functions (PDF $g(r)$--the probability to find a particle pair at distance $r$) at low temperatures (bottom right panel of  Fig.~\ref{fig:EOS}).
%
While in a non-interacting (ideal) classical system, 
$g_{ab}(R)\equiv 1$, interactions and  quantum statistics result in a redistribution of the particles. 
The PDF in Fig.~\ref{fig:EOS} are averaged over the quasiparticle colors. 
The maximum of $r^2 g(r)$ at small interparticle distances supports the interpretation in terms of gluon bound states \cite{Fehske} 
which explains the observed behavior of pressure and density. At the same time, the  quark--quark and 
quark--gluon PDF do not exhibit such peaks and are close to an ideal PDF. 

%
To summarize, in this paper we demonstrated that color quantum 
Monte-Carlo simulations based on the quasiparticle model of the QGP 
are able to reproduce the lattice equation 
of state at zero and non-zero quark chemical potential at
realistic model parameters (quasiparticle masses and coupling constant)
even near and below the critical temperature. Moreover, our color PIMC simulations yield valuable 
insight into the internal structure of the QGP. 
Our results indicate that the QGP exhibits {\em quantum liquid-like} 
properties where the equation of state, quark density and pair distribution 
functions clearly reflect the existence of gluon-gluon bound states (glue balls), 
 at temperatures just below the phase transition, while
meson-like $q\overline{q}$ bound states are not found.  

We acknowledge stimulating discussions with G.~Kalman, P.~Levai, D.~Blaschke, R. Bock and H.~Stoecker.


\end{document}